\documentclass[12pt]{iopart}
\usepackage{graphicx}% Include figure files
\usepackage{iopams}

\begin{document}

\title{Transport Properties of Clean Quantum Point Contacts}

\author{C. R{\"{o}}ssler}
\address{Solid State Physics Laboratory, ETH Zurich, 8093
Zurich, Switzerland} \ead{roessler@phys.ethz.ch}
\author{S. Baer}
\address{Solid State Physics Laboratory, ETH Zurich, 8093
Zurich, Switzerland}
\author{E. de Wiljes}
\address{Solid State Physics Laboratory, ETH Zurich, 8093
Zurich, Switzerland}
\author{P.-L. Ardelt}
\address{Solid State Physics Laboratory, ETH Zurich, 8093
Zurich, Switzerland}
\author{T. Ihn}
\address{Solid State Physics Laboratory, ETH Zurich, 8093
Zurich, Switzerland}
\author{K. Ensslin}
\address{Solid State Physics Laboratory, ETH Zurich, 8093
Zurich, Switzerland}
\author{C. Reichl}
\address{Solid State Physics Laboratory, ETH Zurich, 8093
Zurich, Switzerland}
\author{W. Wegscheider}
\address{Solid State Physics Laboratory, ETH Zurich, 8093
Zurich, Switzerland}

\begin{abstract}
Quantum point contacts are fundamental building blocks for
mesoscopic transport experiments and play an important role in
recent interference- and fractional quantum Hall experiments.
However, it is not clear how electron-electron interactions and the
random disorder potential influence the confinement potential and
give rise to phenomena like the mysterious 0.7 anomaly. Novel growth
techniques of $\rm{Al}_X\rm{Ga}_{1-X}\rm{As}$ heterostructures for
high-mobility two-dimensional electron gases enable us to
investigate quantum point contacts with a strongly suppressed
disorder potential. These clean quantum point contacts indeed show
transport features that are obscured by disorder in standard
samples. From this transport data, we are able to extract the
parameters of the confinement potential which describe its shape in
longitudinal and transverse direction. Knowing the shape (and hence
the slope) of the confinement potential might be crucial to predict
which interaction-induced states can form best in quantum point
contacts.
\end{abstract}

\maketitle

\section{Introduction}
Quantum devices on semiconductor nanostructures rely on quantum
point contacts (QPCs) as basic building blocks. Quantized
conductance has been observed early on~\cite{wha88,wee88} and it has
been used as a signature of the quality of a QPC. With ever
improving sample quality and the perspective for the detection of
non-abelian anyons in the $\nu=5/2$ fractional quantum Hall state,
several experiments~\cite{wil87,mil07,dol08} have recently used the
properties of QPCs fabricated on ultra high-mobility two-dimensional
electron gases (2DEGs). In view of proposals to investigate
fractional quantum Hall states in confined geometries and
interferometer-like setups, the detailed understanding and control
of QPCs are essential. Here we present experimental data which go
beyond previously published data by demonstrating experiments that
profit from the extraordinary cleanliness of the high-mobility 2DEG.
In contrast to standard 2DEGs, we do not observe defect-induced
resonances when the QPCs are shifted laterally. Higher order half
plateaus are observed in the finite-bias differential conductance
(at magnetic field $B_{\perp}=0\,\rm{T}$) as well as spin-split half
plateaus at $B_{\perp}=2\,\rm{T}$. Finally, the $0.7$-anomaly is
investigated as a function of temperature and in perpendicular
magnetic field.

\section{Experiment}
\begin{figure}
\includegraphics[scale=1]{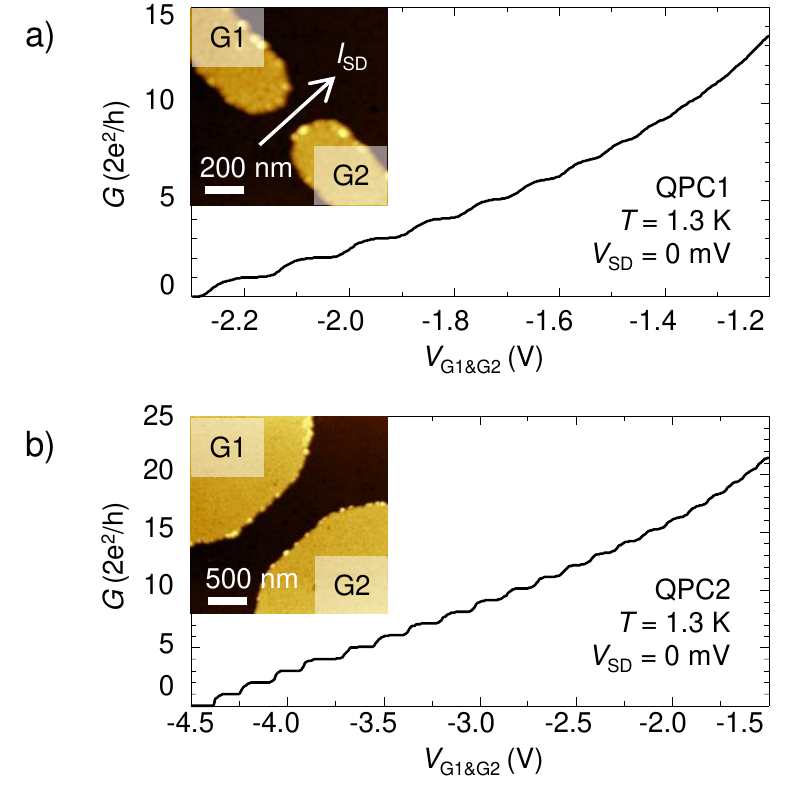}
\caption{\label{fig:sample}a) Inset: Atomic force micrograph of the
sample surface. Two Schottky-gates appear bright, the GaAs surface
appears dark. The distance between the gates is $w=200\,\rm{nm}$.
When the gates are negatively biased, free electrons reside only in
the electron gas underneath the dark areas. Main graph: differential
conductance of QPC1, measured as a function of the voltage applied
to gates G1 and G2. Quantized conductance in multiples of
$G=2\rm{e^2/h}$ indicates the formation of discrete subbands between
the tips of the gates. b) Differential conductance of QPC2, which is
$w=500\,\rm{nm}$ wide and $l\approx1\,\rm{\mu m}$ long.}
\end{figure}

The samples are fabricated on a high-mobility wafer with a
two-dimensional electron gas (2DEG) residing $z=160\,\rm{nm}$
beneath the surface. The high mobility is achieved by placing Si
dopants in a narrow GaAs layer sandwiched by AlAs
layers~\cite{fri96,uma97,hwa08}. The population of the $X$ band in
AlAs results in hardly mobile electrons which screen the static
disorder potential but do not cause a measurable parallel
conductance. Optical lithography is employed to define Hall bars via
mesa etch and deposition of Au/Ge Ohmic contacts. Processed Hall
bars have an electron density of
$n_{\rm{S}}=3.5\times10^{15}\,\rm{m^{-2}}$ and Drude mobilities in
the range of $\mu=1000...2000\,\rm{m^2/Vs}$. The characterization as
well as the following experiments are carried out at a temperature
of $T=1.3\,\rm{K}$, if not stated otherwise. Schottky electrodes are
defined via electron beam lithography and subsequent deposition of
Ti/Au. AFM micrographs of two QPCs are shown in the insets of
Figs.~\ref{fig:sample}a) and b).
The gates appear bright with the gap between them being
$w=200\,\rm{nm}$ (a) and $w=500\,\rm{nm}$ (b). Applying a voltage of
$V_{\rm G}\lesssim-1.1\,\rm{V}$ to the gates depletes the underlying
2DEG and creates a constriction between source and drain. The
source-drain current $I_{\rm{SD}}$ and the voltage drop across the
QPC $V_{\rm{D}}$ are measured in four-terminal configuration while
applying a small lock in amplitude of $V_{\rm{AC}}=100\,\rm{\mu V}$
at a frequency of $f_{\rm{AC}}=33\,\rm{Hz}$ to source and drain. A
DC source-drain voltage $V_{\rm{SD}}$ can be added to $V_{\rm{AC}}$
with both voltages being applied symmetrically with respect to the
common reference potential of source, drain and the gates. Most
transport properties of the employed high-mobility heterostructures
are hysteretic as a function of gate bias~\cite{roe10}. Therefore
all traces are recorded in the same sweep direction by sweeping
towards more negative values of gate voltage.

Figure~\ref{fig:sample}a) shows the differential conductance
$G=dI_{\rm{SD}}/dV_{\rm{D}}$ ($V_{\rm{SD}}=0\,\rm{mV}$) of QPC1,
plotted as a function of the voltage applied to gates G1 and G2.
From the Fermi wavelength of the 2DEG
$\lambda_{\rm{F}}=\sqrt{2\pi/n_{\rm{S}}}=42\,\rm{nm}$ and the
distance of the gates, it would be expected that $n\approx
w/(\lambda_{\rm{F}}/2)=9...10$ modes can be observed due to
confinement transversal to the electron flow~\cite{wee88}. Indeed,
the number of quantized plateaus in Fig.~\ref{fig:sample}a) agrees
with this estimation, indicating that the largest electronic width
of the QPC matches the lithographic gap of the Schottky split-gates.
Due to the larger gate-spacing of QPC2, correspondingly more modes
are observed in Fig.~\ref{fig:sample}b) and a significantly larger
gate bias has to be applied in order to pinch off. We observe
irreversible charging of the sample typically at
$V_{\rm{G}}\sim-5\,\rm{V}$ which limits the QPCs' range of
operation.

\subsection{Lateral Shifting of the QPC}

\begin{figure}
\includegraphics[scale=1]{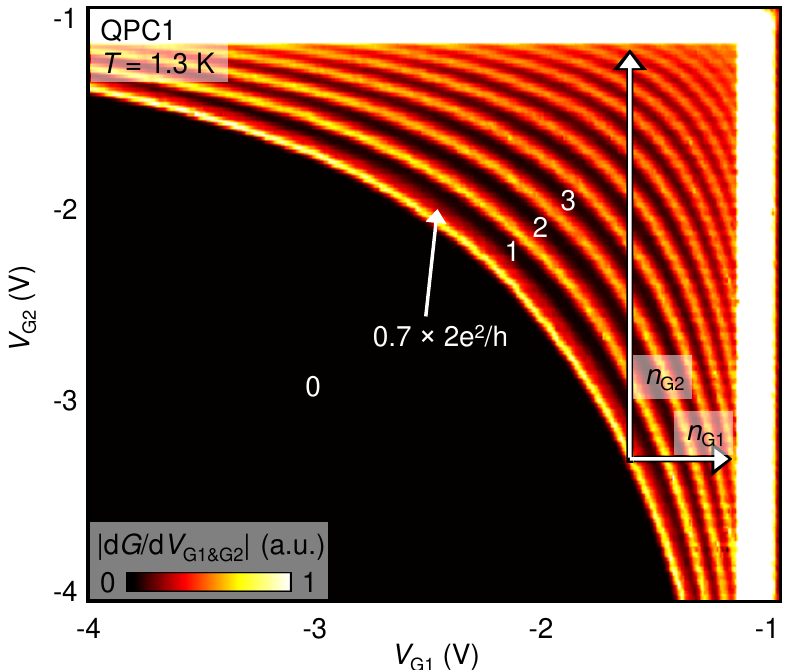}
\caption{\label{fig:scanning} (color online) Transconductance
$G_{\rm{TC}}=dG/dV_{\rm{G1\&G2}}$ of QPC1, plotted in false colors
as a function of the voltages applied to gates G1 and G2. Integer
conductance values of $G=0,1,2,3,...\times\rm{2e^2/h}$ result in
$G_{\rm{TC}}=0$ (black), steps in-between integer conductance values
appear bright. A faint red stripe with $G\approx0.7\times\rm{2e^2/h}$
indicates the presence of a $0.7$-anomaly in the QPC. The strong
increase of the conductance at the right and top border (white)
marks the gate-pinch-off, where the 2DEG underneath gates G1 (right)
and G2 (top) starts to be depleted. If both gates are biased
identically, 12-13 plateaus are observed between QPC-pinch-off and
gate-pinch-off. When the ratio of gate-voltages is varied by
following the lowest transconductance-stripe, the number of plateaus
in dependence of either $V_{\rm{G1}}$ or $V_{\rm{G2}}$ can be
varied, indicating that the QPC is shifted laterally between the
gates. Scattering centers in-between the gates would appear as
straight lines with the slope corresponding to their capacitance to
gates G1 and G2. No such defects are visible in this scan.}
\end{figure}

QPC1 can be further characterized by varying the voltages applied to
each of the gates, which is not possible for QPC2 due to its extreme
pinch-off voltage. Figure~\ref{fig:scanning} shows the
transconductance $G_{\rm{TC}}=dG/dV_{\rm{G1\&G2}}$ of QPC1 in
grayscale, plotted as a function of $V_{\rm{G1}}$ and $V_{\rm{G2}}$.
Black areas correspond to pinch-off (bottom left) and successive
conductance plateaus (marked by $1,2,3$). Such a plot reveals
scattering centers in the channel, since changing the ratio of gate
voltages causes the position of the channel to shift laterally
between the gates~\cite{wil90}. The shift can be approximately
determined by counting the number of steps ($n_{\rm{G1}}$,
$n_{\rm{G2}}$) that can be observed as a function of each
gate~\cite{sch11}: $\Delta
y=\lambda_{\rm{F}}/2\times(n_{\rm{G1}}-n_{\rm{G2}})/2$. In the
situation marked by white arrows in Fig.~\ref{fig:scanning} this
amounts to $\Delta y=21\,\rm{nm}\times(9-4)/2=53\,\rm{nm}$. The
largest observed shift is $\Delta
y=\pm21\,\rm{nm}\times(11-2)/2=\pm95\,\rm{nm}$, which corresponds to
the lithographic distance of the gates. When the QPC is shifted from
one gate to the other, its potential would naturally change if
localized impurities in the channel or the static disorder potential
created by the dopants were relevant. Since these defects are fixed
in space, they should appear as lines intersecting the QPC steps,
with their slope given by the capacitance to gates G1 and G2. The
absence of such defect-induced lines confirms the cleanliness of the
sample and the effectiveness of the screening layers in suppressing
the charged dopants' disorder potential.

\subsection{The 0.7-Anomaly in Perpendicular Magnetic Field}

\begin{figure}
\includegraphics[scale=1]{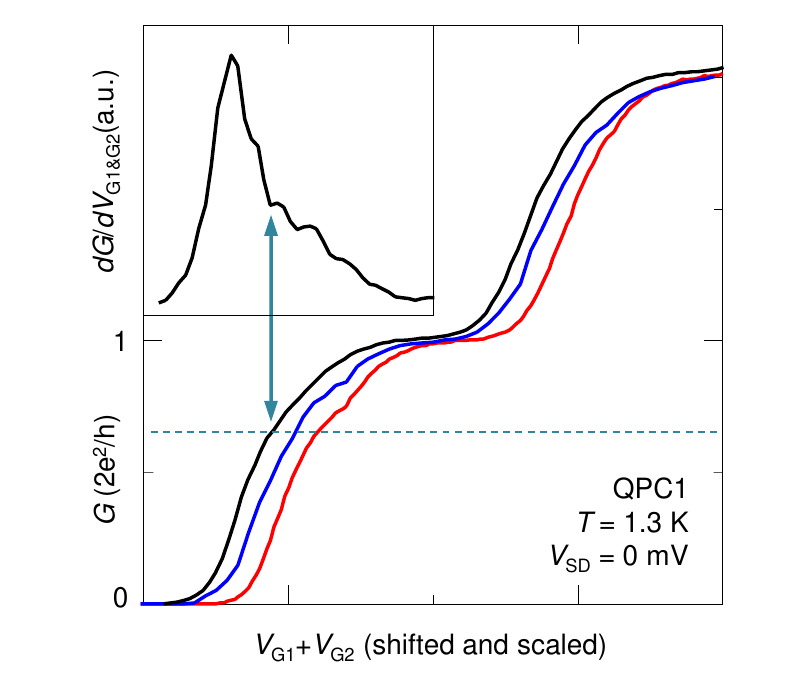}
\caption{\label{fig:0.7shift} (color online) Pinch-off traces for
different positions of the QPC. The traces have been shifted and
linearly scaled along the gate-axis for better comparison. From left
to right: simultaneously sweeping $V_{\rm{G1}}=V_{\rm{G2}}$;
sweeping $V_{\rm{G2}}$ while $V_{\rm{G1}}=-1.7\,\rm{V}$; sweeping
$V_{\rm{G1}}$ while $V_{\rm{G2}}=-1.7\,\rm{V}$. The 0.7-anomaly
appears as a shoulder (change of slope $dG/dV_{\rm{G1\&G2}}$ - see
inset) at $G\approx0.65\times2\rm{e^2/h}$. Switching events in the
two rightmost traces arise because the more negative gate voltage
has to be biased with $V_{\rm{G}}\lesssim-3\,\rm{V}$ in order to
pinch off.}
\end{figure}

It is noteworthy that the 0.7 anomaly~\cite{tho96,kri00,cro02}, an
additional plateau with a conductance of
$G\approx0.7\times\rm{2e^2/h}$, appears as a weak shoulder close to
pinch-off of both QPCs. In the transconductance-plot in
Fig.~\ref{fig:scanning}, the 0.7 anomaly is visible as an asymmetry
of the pinch-off line, giving rise to a gray (red) stripe adjacent
to the $G=1\times\rm{2e^2/h}$ plateau. The 0.7-stripe is continuous
and reaches all the way to the extreme QPC shifts, emphasizing that
the 0.7-anomaly is an intrinsic property of the QPC. Cuts along the
diagonal (where $V_{\rm{G1}}=V_{\rm{G2}}$ as well as strongly
shifted configurations (either $V_{\rm{G1}}$ or $V_{\rm{G2}}$ being
fixed at $-1.7\,\rm{V}$) are shown in Fig.~\ref{fig:0.7shift}.
When the QPC is defined centrally in-between the gates
($V_{\rm{G1}}=V_{\rm{G2}}$, leftmost trace), the 0.7-anomaly
manifests itself as a weak shoulder below the
$G=2\rm{e^2/h}$-plateau. The top left inset shows the numerically
derived slope $dG/dV_{\rm{G1\&G2}}$ which exhibits a clear change of
slope at the position marked by an arrow. The corresponding
conductance at this gate voltage is $G=0.65\times2\rm{e^2/h}$. For
comparison, two configurations with the QPC being defined close to
gate G2 (central trace) or gate G1 (rightmost trace) are plotted on
the same gate axis. We find that both asymmetrically measured traces
resemble the shape of the symmetric case. The conductance value of
the 0.7-anomaly does not change when shifting the QPC laterally,
however our accuracy of determining it is limited to
$G=0.65\pm0.05\times2\rm{e^2/h}$ due to switching events caused by
the more negative gate voltages required for pinch-off in an
asymmetric gate configuration. In agreement with previous
studies~\cite{tho96,kri00,cro02}, we find that the 0.7-anomaly is
less pronounced in 2DEGs with high density (here:
$n_{\rm{S}}=3.5\times10^{15}\,\rm{m^{-2}}$) compared to samples with
electron densities in the range of
$n_{\rm{S}}\sim1\times10^{15}\,\rm{m^{-2}}$.

\begin{figure}
\includegraphics[scale=1]{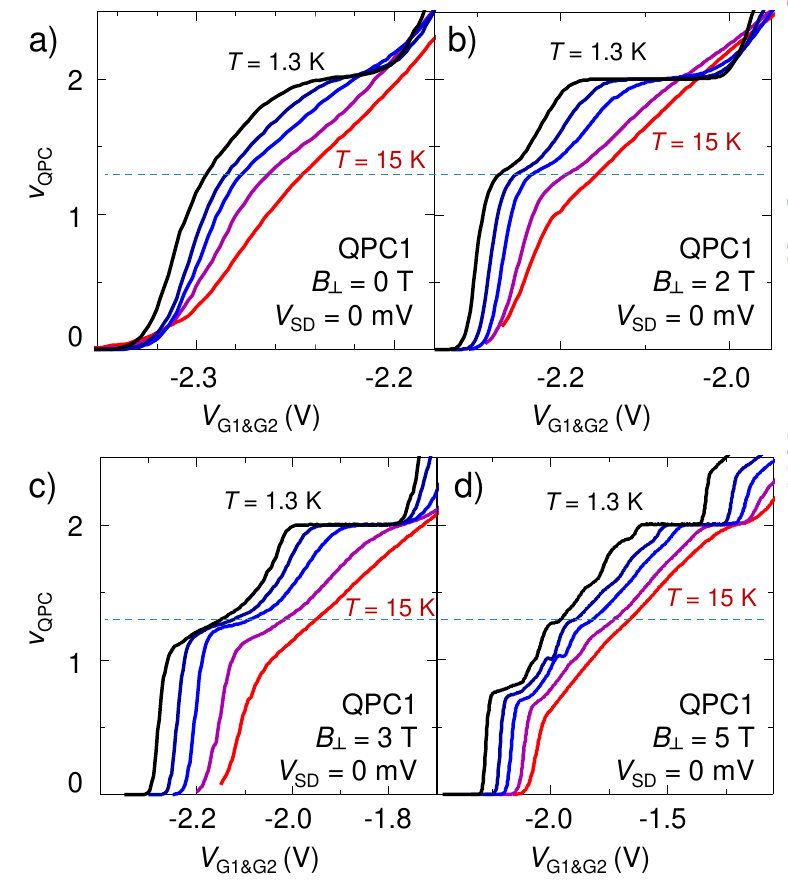}
\caption{\label{fig:0.7-t-b} (color online) Transmission
$\nu_{\rm{QPC}}$ through QPC1 measured as a function of
symmetrically applied gate bias at different temperatures (from left
to right:
$T=1.3\,\rm{K}/2.5\,\rm{K}/5\,\rm{K}/10\,\rm{K}/15\,\rm{K}$). The
traces are plotted with horizontal offset for clarity. A dashed
horizontal line marks the transmission value of $\nu_{\rm{QPC}}=1.3$
associated with the 0.7-anomaly. a) The 0.7-anomaly becomes more
pronounced with increasing temperature. At $T=10\,\rm{K}$ the
subband quantization is completely washed out but the 0.7-anomaly is
still clearly visible. b) In a magnetic field of
$B_{\perp}=2\,\rm{T}$ applied perpendicular to the plane of the
2DEG, the shoulder at $\nu_{\rm{QPC}}=1.3$ is well developed at low
temperature and shifts to lower transmission with increased
temperature. c) At $B_{\perp}=3\,\rm{T}$, the transmission shows a
non-monotonic behavior as a function of temperature. The
transmission of the 0.7-anomaly first recovers almost to its
zero-field value, then decreases again for $T>5\,\rm{K}$. d) At
$B_{\perp}=5\,\rm{T}$, various plateaus related to the transmission
of (fractional) edge channels are observed. These features wash out
when the temperature is increased.}
\end{figure}

In order to compare the results obtained on low-density 2DEGs to the
behavior of our system, the temperature- and
magnetic-field-dependence is investigated in detail. However,
similar to recent work performed on comparable high-mobility
2DEGs~\cite{hat11}, we find a suppression of the Hall mobility in
parallel magnetic field which is in our devices accompanied by a
suppression of the QPCs' spin splitting (data not shown). We are
hence limited to applying a magnetic field perpendicular to the
2DEG, which should also weaken the 0.7-anomaly by lifting the spin
degeneracy. Since the differential conductance is strongly modified
by the presence of edge channels in the quantum Hall regime, the
filling factor $\nu_{\rm{QPC}}$ is obtained from the diagonal
voltage drop across QPC1. Figure~\ref{fig:0.7-t-b}a) shows the lower
part of the pinch-off trace for different temperatures without a
magnetic field being applied.
As expected, the 0.7-anomaly evolves into a more pronounced shoulder
when the temperature is increased from $T=1.3\,\rm{K}$ (left) to
$T=15\,\rm{K}$ (right). The marked value of $\nu_{\rm{QPC}}=1.3$
(dashed line) corresponds to the conductance of
$G=0.65\times2\rm{e^2/h}$ extracted from Fig.~\ref{fig:0.7shift}.

By applying a magnetic field perpendicular to the 2DEG, we expect
the 0.7-anomaly to be influenced by the increase of both the
energetic and spatial separation of the lowest two spin channels.
This idea of ``mimicking the 0.7 scenario'' was previously
investigated in \cite{sha06}, but the interpretation of the data
proved difficult due to additional resonances in the pinch-off
traces. Figure~\ref{fig:0.7-t-b}b) is recorded at
$B_{\perp}=2\,\rm{T}$, where spin-resolved edge channels begin to
form at $T=1.3\,\rm{K}$. The data shows a well pronounced plateau at
$\nu_{\rm{QPC}}=1.3$ which is weakened when the
temperature is increased to $T>5\,\rm{K}$. Figure~\ref{fig:0.7-t-b}c) shows the
same measurement at $B_{\perp}=3\,\rm{T}$, where the edge channels
are further separated energetically as well as spatially. Now, the
temperature dependence is non-monotonic with the 0.7-anomaly first
rising almost to the expected transmission of $\nu_{\rm{QPC}}=1.3$,
then decaying to lower transmission. Data taken at
$B_{\perp}=5\,\rm{T}$ is shown in Fig.~\ref{fig:0.7-t-b}d). The
structure is more complex now due to the formation of fractional
edge channels and does not show a feature which is unambiguously
related to the 0.7-anomaly. The observed shoulders and plateaus wash
out, perhaps with the plateau at $\nu_{\rm{QPC}}\approx0.7$ being
more resilient than all other features at $\nu_{\rm{QPC}}<2$. The
overall dependence of the 0.7-anomaly on magnetic field is in
contrast to observations in two-dimensional hole gases where the
0.7-anomaly was found to evolve into a resonance for strong
perpendicular magnetic field~\cite{kom10}. There, the appearance of
a resonance was discussed in view of a quasi-localized state in
combination with Kondo effect.

One possible interpretation of our magnetic field dependence follows
the idea of two spin-polarized channels leading to the 0.7
scenario~\cite{sha06}. Moderate magnetic field
($B_{\perp}=2\,\rm{T}$) increases the spin polarization, thereby
enhancing the 0.7-anomaly. Stronger fields increase the spatial
separation between the edge channels, thereby reducing interactions
and weakening the 0.7-anomaly. At $B_{\perp}=3\,\rm{T}$, the spatial
separation can be overcome by increasing the temperature to a value
where thermal energy and B-field-induced spin splitting become
comparable in magnitude. At even stronger magnetic field, other
many-body effects besides the 0.7-anomaly might become relevant
which makes a detailed interpretation difficult. For future studies
it might prove worthwhile to investigate the zero-bias anomaly in
perpendicular magnetic field in order to check if the interpretation
of spatially separated edge channels is consistent with other
experimental findings.

\subsection{Finite Bias Spectroscopy}

\begin{figure}
\includegraphics[scale=1]{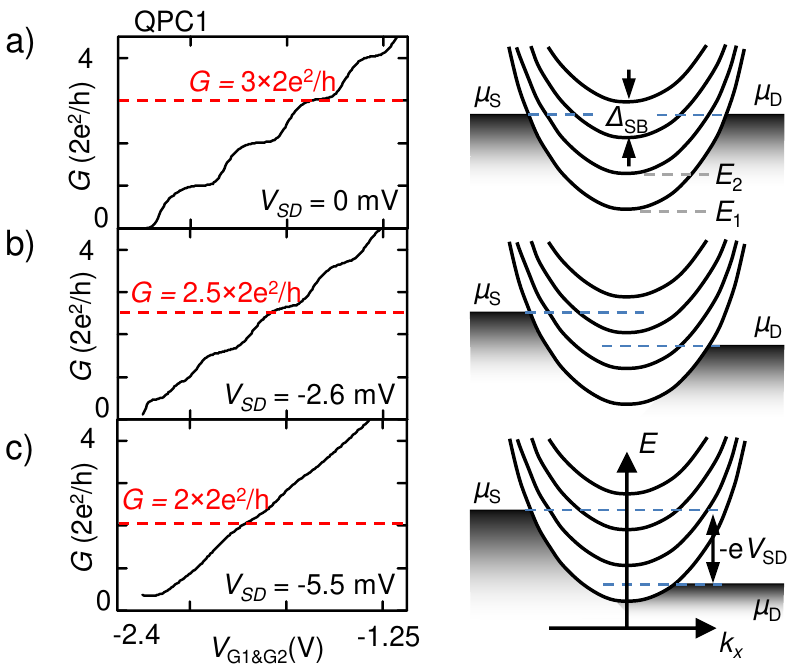}
\caption{\label{fig:bias} (color online) Pinch-off trace (left) and
schematic view (right) for different values of the source-drain bias
$V_{\rm{SD}}$. a) Linear-response regime $V_{\rm{SD}}=0\,\rm{mV}$.
The measured trace displays plateaus at the expected conductance
values. The conductance value of $G=3\times\,\rm{2e^2/h}$
corresponding to the sketched situation (right) is marked by a
dashed line. In the depicted situation the gate voltage is set such
that three (spin degenerate) subband bottoms lie below the chemical
potential of source ($\mu_{\rm{S}}$) and drain ($\mu_{\rm{D}}$). The
energies of the subband-bottoms are labeled $E_1$, $E_2$,... b)
Finite bias measurement with one subband bottom in-between
$\mu_{\rm{S}}$ and $\mu_{\rm{D}}$. The expected half-plateau
conductance of $G=2.5\times\,\rm{2e^2/h}$ is marked by a dashed line
in the experimental trace. c) Two subband bottoms reside in-between
$\mu_{\rm{S}}$ and $\mu_{\rm{D}}$. Integer plateau values are
expected and can be observed as shoulders in the pinch-off curve.}
\end{figure}

Further characterization of QPC1 requires finite-bias measurements,
because employing $V_{\rm{SD}}$ as an energy reference gives access
to the QPC's subband spacings~\cite{gla89,pat90,pat91,tho95}. Three
exemplary gate traces are depicted in Fig.~\ref{fig:bias}.
Figure~\ref{fig:bias}a) shows the linear-response regime
$V_{\rm{SD}}=0\,\rm{mV}$ which is identical to the trace shown in
Fig.~\ref{fig:sample}a). A sketch of the energy landscape is shown
at the right hand side. The parabolic electron dispersions are
energetically separated by the subband spacing $\Delta_{\rm{SB}}$
due to transversal confinement. In the depicted situation, three
subband-bottoms reside below the chemical potentials of source and
drain, giving rise to a conductance of $G=3\times\,\rm{2e^2/h}$.
Figure~\ref{fig:bias}b) shows the pinch-off trace for
$V_{\rm{SD}}=-2.6\,\rm{mV}$, where plateaus appear at half-integer
values of the conductance. The sketch corresponding to a conductance
of $G=2.5\times\,\rm{2e^2/h}$ is shown on the right hand side: two
subbands contribute fully and one subband contributes half to the
overall conductance.

So-called half-plateaus can only be observed in clean samples,
presumably because scattering events inside the QPC become more
likely when more unoccupied subband states are energetically
available at larger source-drain bias. At even higher bias, the
conductance is usually obscured by noise~\cite{roe10b} or it
increases/decreases due to various self-gating
effects~\cite{kri00,pat91}. In QPC1 however, the return of integer
conductance quantization for two subband bottoms residing in-between
$\mu_{\rm S}$ and $\mu_{\rm D}$ is observable in
Fig.~\ref{fig:bias}c) at $V_{\rm{SD}}=-5.5\,\rm{mV}$. We interpret
this observation as another result of the cleanliness of the QPC
which reduces the probability for backscattering.

\begin{figure}
\includegraphics[scale=1]{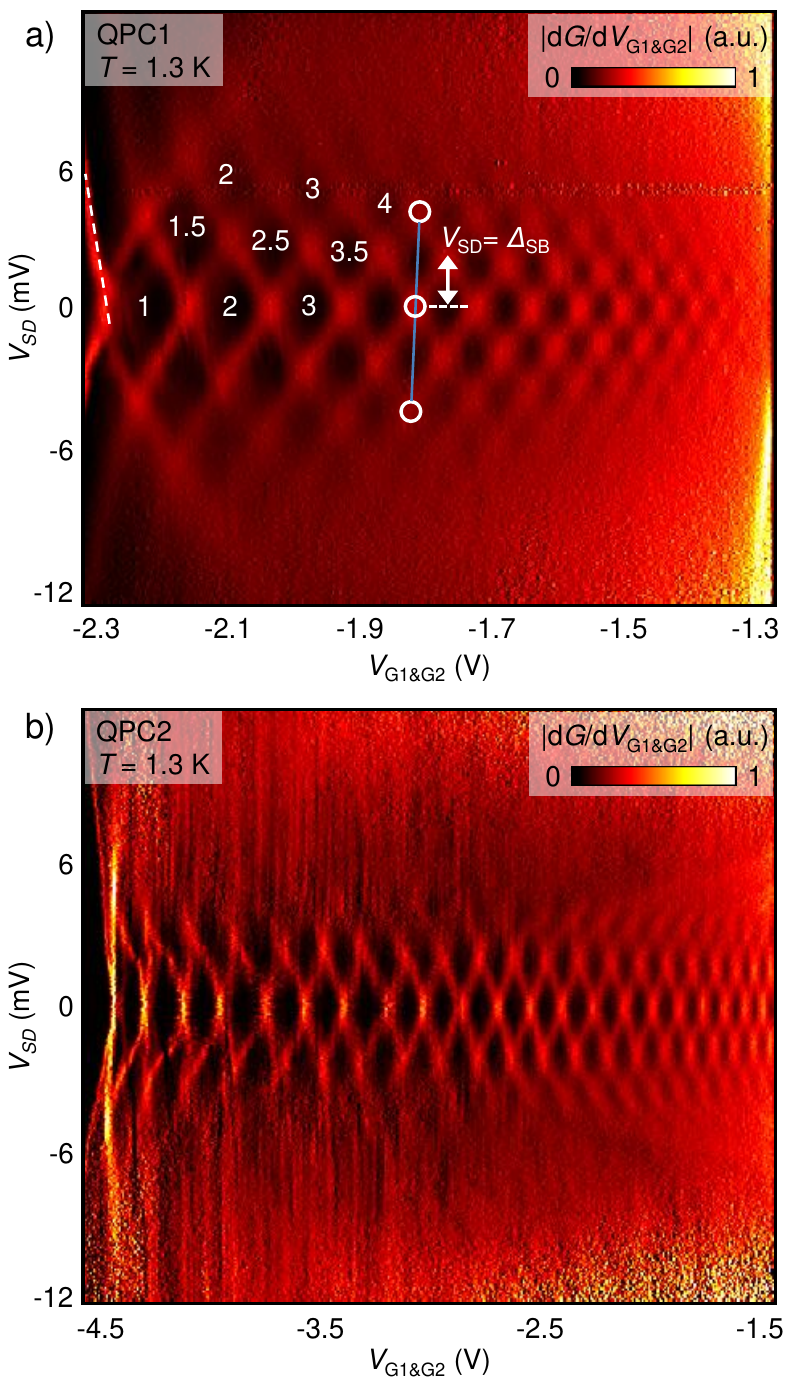}
\caption{\label{fig:nonlinear} (color online) a) Transconductance
$G_{\rm{TC}}=dG/dV_{\rm{G1\&G2}}$ of QPC1, plotted as a function of
source-drain bias $V_{\rm{SD}}$ and voltage $V_{\rm{G1\&G2}}$
applied to gates G1 and G2. Plateaus in the differential conductance
of $G=1,2,3,...\times\,\rm{2e^2/h}$ appear as black diamonds
centered around $V_{\rm{SD}}=0\,\rm{mV}$. Their extent in
$V_{\rm{SD}}$ corresponds to the subband-spacing $\Delta_{\rm{SB}}$.
Higher order half-plateaus ($G=1.5,2.5,3.5,...\times\,\rm{2e^2/h}$)
and second-order integer-plateaus ($G=2,3,4,...\times\,\rm{2e^2/h}$)
appear as black diamonds at finite source-drain bias. b)
Transconductance of QPC2. Integer plateaus around
$V_{\rm{SD}}=0\,\rm{mV}$ are resolved. For gate voltages
$V_{\rm{G1\&G2}}\lesssim-2.5\,\rm{V}$ higher order plateaus are
obscured by noise.}
\end{figure}

In order to retrieve the full information about the confinement
potential, the transconductance of QPC1 is plotted in
Fig.~\ref{fig:nonlinear}a) as a function of $V_{\rm{SD}}$ and
$V_{\rm{G1\&G2}}$.
Integer conductance plateaus without subband minima between
$\mu_{\rm S}$ and $\mu_{\rm D}$ (labeled $1,2,3$), appear as black
diamonds around $V_{\rm{SD}}=0\,\rm{mV}$. Half plateaus
($1.5,2.5,3.5$) and second order integer plateaus ($2,3,4$) appear
in a regular pattern at finite source-drain bias.

In our experience, the higher order plateaus can not be observed in
samples with mobility $\mu\lesssim10\,\rm{m^2/Vs}$
(cf.~\cite{roe10b}). Comparing our data to data of a defect-free
QPC~\cite{tho95} defined in a 2DEG with mobility
$\mu=150\,\rm{m^2/Vs}$, we find subtle differences in the
transconductance pattern. Although the QPC in reference~\cite{tho95}
is measured at a lower temperature ($T=90\,\rm{mK}$) than our device
($T=1.3\,\rm{K}$), second order integer plateaus seem to be
suppressed as long as no magnetic field is applied perpendicular to
the 2DEG. The authors state that ``Due to the suppression of
backscattering in the presence of a small magnetic field the
reappearance of the integer plateaus at high $V_{\rm{SD}}$ can be
clearly observed''. Our QPC is defined in a 2DEG with mobility
$\mu\gtrsim1000\,\rm{m^2/Vs}$ and second order integer plateaus are
clearly resolved at $B_{\perp}=0\,\rm{T}$. These observations
indicate that even though the mean free path of the electrons is
much larger than the length of the QPC in both cases, a higher
electron mobility still manifests itself as reduced backscattering
in the regime of nonlinear conductance.

As seen from the sketches in Fig.~\ref{fig:bias}, the maximum extent
of the diamonds in $V_{\rm{SD}}$ corresponds to the energy spacing
$\Delta_{\rm{SB}}$ of the involved QPC subbands. The whole pattern
of transconductance diamonds is sheared with features at positive
$V_{\rm{SD}}$ shifted by about $5\,\%$ to more positive gate voltage
than in a perfectly symmetric configuration. This asymmetry could
hint at a slight asymmetry of the QPC's coupling to source and drain
but might also be explained by a gradual drift of the local
potential over the measurement time of $21\,\rm{hours}$.

The effect of $V_{\rm{SD}}$ onto the confinement potential, the
so-called self-gating, manifests itself as a deviation from a
pattern of straight lines~\cite{kri00}. From
Fig.~\ref{fig:nonlinear}a) it appears that self-gating plays an
important role mainly close to pinch-off (white dashed line) and
perhaps at very large $V_{\rm{SD}}$, where no clear quantization is
observed any more. Since self-gating appears not to dominate the
shape of the transconductance pattern it is possible to learn more
about the confinement potential by comparing the position of
transconductance nodes in Fig.~\ref{fig:nonlinear}a). Three
exemplary nodes are highlighted by white circles. They correspond to
the resonance conditions (from top to bottom) ($\mu_{\rm
S}=E_4,\mu_{\rm D}=E_6$), ($\mu_{\rm S}=E_5=\mu_{\rm D}$) and
($\mu_{\rm S}=E_6,\mu_{\rm D}=E_4$), respectively. The fact that
these resonance conditions occur at almost the same gate voltage
(along a straight line) means that the subband-spacings
$\Delta_{45}=E_5-E_4$ and $\Delta_{56}=E_6-E_5$ are very similar at
this gate voltage (compare sketches in Fig.~\ref{fig:bias}).
Therefore the transversal confinement can be well described by a
harmonic potential. If the confinement potential were for example a
square well, the subband spacings would increase with higher mode
number and hence the higher order modes would occur at more positive
gate voltage than the linear response node.

As a comparison to the clean and regular pattern of QPC1,
Figure~\ref{fig:nonlinear}b) shows the transconductance of QPC2.
Higher order plateaus are visible for
$V_{\rm{G1\&G2}}\gtrsim-2.5\,\rm{V}$ but are obscured by noise at
more negative bias which is usually related to tunneling events from
the gates into the doping layer~\cite{pio05}. Furthermore, the shape
of the five leftmost diamonds is distorted with the upper and lower
tip being shifted to more negative gate bias. This shift as well as
the curvature of the plateau borders follows the dependence observed
in quantum wires and arises from the requirement of satisfying
charge neutrality with a 1-dimensional density of states while
applying a finite source-drain voltage~\cite{pic04}. The quantum
wire-like characteristic is consistent with the geometry of the
gates which should create a channel that is longer than the
screening length of the 2DEG (compare inset of
Fig.~\ref{fig:sample}b)). It is noteworthy that also in the quantum
wire-like QPC2, we observe a well pronounced half-plateau related to
the 0.7-anomaly which resembles the features observed
in~\cite{pic04}. Since we do not observe defect-related resonances
in QPC2, the design might be extended to even longer gate-defined
quantum wires~\cite{tar95,lia99,mor06} in order to study the
length-dependence of the 0.7-anomaly. However, the observation of
diffusive transport~\cite{lia99} in $L_{\rm{QWR}}\geq5\,\rm{\mu m}$
long quantum wires (with the mean free path in the 2DEG being
$L_{\rm{MF}}\sim40\,\rm{\mu m}$) suggests that gate-defined quantum
wires might not profit from an increased free electron mobility at
least if split-gate technology is used for confinement.

\subsection{Extracting the QPCs' Shape Parameters}

\begin{figure}
\includegraphics[scale=1]{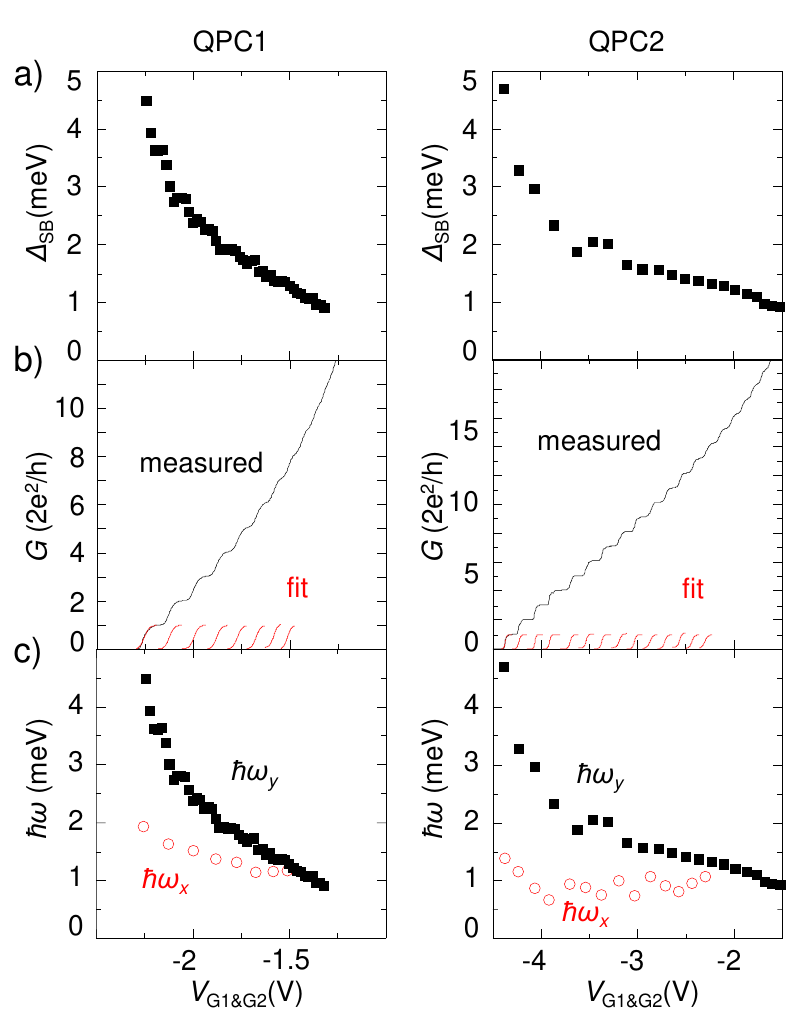}
\caption{\label{fig:fits} (color online) a) Subband spacings
$\Delta_{\rm{SB}}$ of QPC1 (left) and QPC2 (right), as determined
from finite bias transport measurements. With more negative gate
voltage, the subband spacings increase. b) Differential conductance,
plotted as a function of gate voltage. The measured curve (black)
can be fit by assuming a saddle-point potential and calculating the
transversal harmonic confinement potential from the subband spacing
in a). The resulting fit for each step is plotted at the bottom. c)
Subband spacing of the transversal (squares) and longitudinal
(circles) confinement potential, as extracted from the fits in b).
Approaching pinch-off, QPC1 (left) becomes much narrower but only
slightly shorter. The longitudinal curvature of QPC2 (right) is
smaller than that of QPC1, which is in agreement with the
lithographic dimensions of the gates.}
\end{figure}

As discussed earlier, the transconductance plot can now be used to
reconstruct the confinement potential. The subband spacing can be
determined from the $V_{\rm{SD}}$ position of the borders of the
transconductance diamonds~\cite{pat91}. Due to the finite resistance
of the leads $R_{\rm{S}}=400\,\rm{\Omega}$, a fraction of the
applied DC-bias $V_{\rm{SD}}$ does not drop at the QPC. Using
$R_{\rm{S}}$ and the measured four-terminal conductance $G$, this is
taken into account via
$\Delta_{\rm{SB}}=V_{\rm{SD}}/(1+GR_{\rm{S}})$. The thereby
determined subband spacings are plotted as a function of
$V_{\rm{G1\&G2}}$ in Fig.~\ref{fig:fits}a) for QPC1 (left) and QPC2
(right).
The subband spacings increase monotonically with more negative gate
bias, indicating that the confinement potential becomes narrower and
steeper while approaching pinch-off. This trend has been observed
before~\cite{pat91} and can be explained by the reduced influence of
screening on the confinement potential when the local electron
density is reduced~\cite{lau88}. Since the higher order plateaus
indicate that the confinement potential of QPC1 has a close-to
harmonic shape, we can now use the measured subband spacings to
apply B{\"{u}}ttiker's saddle-point model~\cite{bue90} to our linear
response data and extract all parameters of the potential profile at
the constriction. Since QPC2 shows quantum wire-like transport
characteristics, the model is not expected to reflect the exact
potential shape of QPC2, but should still give qualitatively
meaningful results. Temperature broadening is not accounted for in
this model since the subband-spacings $\Delta_{\rm{SB}}>1\,\rm{meV}$
are much larger than the thermal broadening
$k_{\rm{B}}T\approx0.1\rm{meV}$.

Neglecting inter-mode scattering and including spin degeneracy, the
transmission of the $n$-th subband is given by
\begin{equation}
T_n=2/(1+exp(-\pi\varepsilon_n/\hbar\omega_X))
    \label{eq:transmission}
\end{equation}
with the energy of the $n$-th subband
\begin{equation}
\varepsilon_n=2(\hbar\omega_Y\,(n+1/2)-E_{\rm{CB}}).
    \label{eq:epsilon}
\end{equation}
The gate dependence $\omega_Y(V_{\rm{G1\&G2}})$ is known from
Fig.\ref{fig:fits}a), the gate dependence of the conduction band
bottom $E_{\rm{ CB}}(V_{\rm{G1\&G2}})$ is approximated by
$E_{\rm{CB}}(V_{\rm{G1\&G2}})=E_0+\alpha\times V_{\rm{G1\&G2}}$ for
each conductance step with the lever arm $\alpha$ converting gate
voltage to energy. Usually, the lever arm can be determined by
taking the source-drain bias as an energy reference and comparing it
to the gate dependence of a given transport resonance~\cite{kou01}.
But as shown in Fig.\ref{fig:fits}a), the gate voltage not only
lifts $E_{\rm{CB}}$ but also increases the subband spacing, giving
rise to a seemingly increased lever arm. Knowing that the
confinement potential is harmonic enables an alternative way to
determine $\alpha$. At the position of the conductance steps in
Fig.\ref{fig:fits}a) (at $G=(n-0.5)\times\rm{2e^2/h}$), the
conduction band bottom is $E_{\rm{CB}}=(n-0.5)\,\hbar\omega_Y$ below
the Fermi energy. Hence, the lever arm in-between two successive
steps is given by
\begin{equation}
\alpha=(n+1-1/2)\,\hbar\omega_{Y,n+1}-(n-1/2)\,\hbar\omega_{Y,n}
    \label{eq:lever}
\end{equation}
with $\omega_{Y,n}$ being the confinement at the $n$-th step as
extracted from Fig.\ref{fig:fits}a). Now the only fitting parameters
are the longitudinal curvature $\omega_X$ and the energy offset
$E_0$.

Figure~\ref{fig:fits}b) shows the pinch off curve of QPC1 (left) and
the fits resulting from the described procedure. All fits are
plotted below $G={2e^2/h}$ for clarity. The same procedure applied
to QPC2 is shown on the right hand side. Figure~\ref{fig:fits}c)
shows the extracted values for the longitudinal curvature (circles)
of QPC1 (left) and QPC2 (right). The transversal curvature is
plotted as squares for comparison. In both QPCs, $\omega_X$ is
smaller than $\omega_Y$, as required for the observation of
conductance quantization~\cite{bue90}. Approaching pinch-off, QPC1
becomes much narrower (strong increase of $\hbar\omega_Y$) and
slightly shorter (increase of $\hbar\omega_X$). QPC2 also becomes
much narrower but there is no strong increase of $\hbar\omega_X$.
Although the saddle-point model might not be the ideal model for a
quantum wire, this fits the intuitive picture of a 1-dimensional
channel with a lithography-defined length and voltage-controlled
width.

Comparing the parameters obtained from our analysis to those from
earlier investigations, we find surprising discrepancies despite
similar 2DEG density and gate-spacings. The data analyzed
in~\cite{tab95} is well described by $\hbar\omega_Y=0.9\,\rm{meV}$,
$\hbar\omega_X=0.3\,\rm{meV}$ with both values being independent of
gate voltage. In our devices, the shape of the confinement potential
changes dramatically as a function of gate voltage and reaches
oscillator strengths of $\hbar\omega_Y>4\,\rm{meV}$,
$\hbar\omega_X>1\,\rm{meV}$ close to pinch-off. We interpret this
observation as the result of the screening properties of the
screening layers that are grown into the heterostructure in order to
achieve ultra-high electron mobilities~\cite{fri96,uma97,hwa08}.
Screening should reduce the range of the gate-induced potential and
thereby increase the slope of the confinement potential.

\subsection{Spin-Resolved Transport at Low Temperatures}

\begin{figure}
\includegraphics[scale=1]{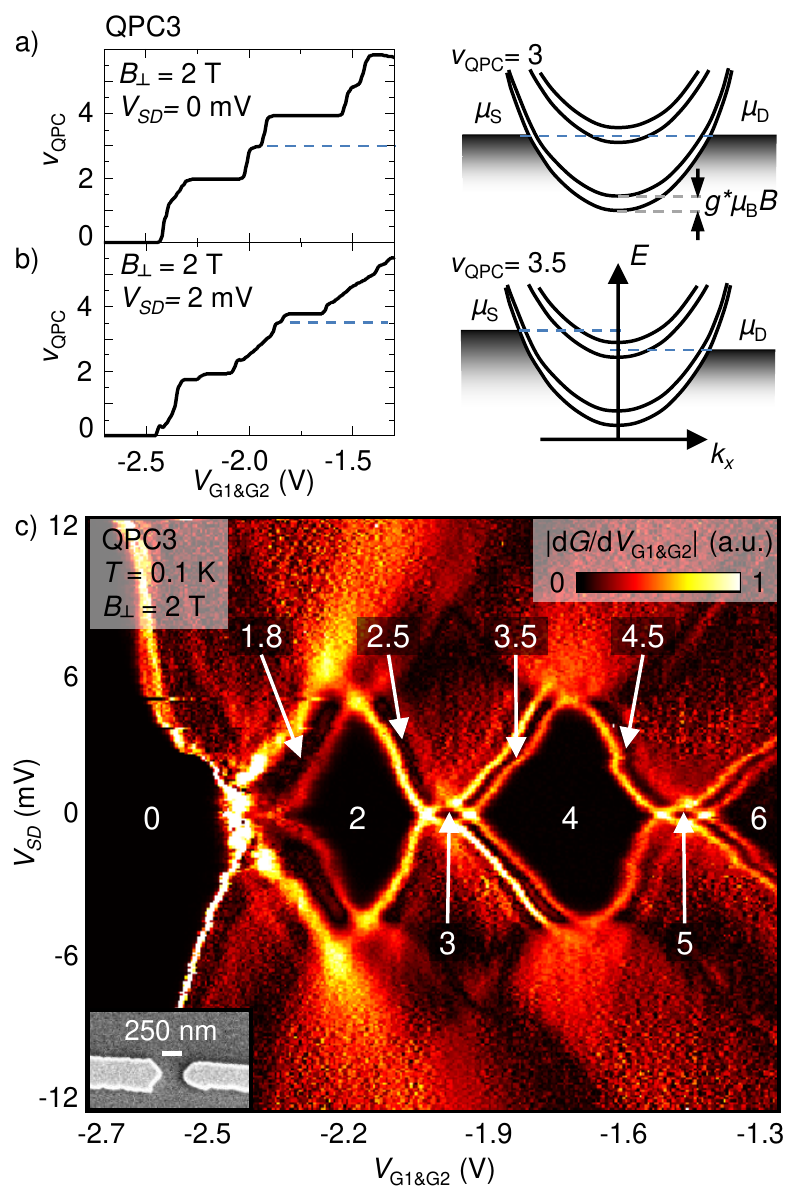}
\caption{\label{fig:bfield} (color online) a) Filling factor
$\nu_{\rm{QPC}}$ of QPC3 (split-gate gap $w=250\,\rm{nm}$), plotted
as a function of $V_{\rm{G1\&G2}}$. Measured at a temperature of
$T=100\,\rm{mK}$, the spin degeneracy is lifted by a magnetic field
$B_{\perp}=2\,\rm{T}$ applied perpendicular to the plane of the
2DEG. The situation corresponding to $\nu_{\rm{QPC}}=3$ is sketched
on the right hand side: three non-degenerate subbands reside below
the chemical potentials of source and drain. b) Filling factor of
QPC3 as a function of $V_{\rm{G1\&G2}}$ while a source-drain bias of
$V_{\rm{SD}}=2\,\rm{mV}$ is applied. Filling factors
$\nu_{\rm{QPC}}=2.5$ and $\nu_{\rm{QPC}}=3.5$ are observed when one
spin-split level lies in-between $\mu_{\rm{S}}$ and $\mu_{\rm{D}}$
(sketched on the right hand side). c) False color plot of the
transconductance $G_{\rm{TC}}=dG/dV_{\rm{G1\&G2}}$ of QPC3 (inset),
plotted as a function of $V_{\rm{SD}}$ and $V_{\rm{G1\&G2}}$.
Regions of integer filling factor $\nu_{\rm{QPC}}=2,3,4,5,6$ appear
as black diamonds centered around $V_{\rm{SD}}=0\,\rm{mV}$, half
plateaus with $\nu_{\rm{QPC}}=2.5,3.5,4.5$ appear as black stripes
at finite source-drain bias. The 0.7 anomaly creates a plateau with
$\nu_{\rm{QPC}}=1.8$.}
\end{figure}

Additional changes in the confinement can be created by applying a
magnetic field $B_{\perp}$ perpendicular to the plane of the 2DEG
which lifts the spin degeneracy and increases the subband
spacing~\cite{wha88,wee88b,tho95,tho96,roe10}. QPC3 was not equipped
with a 2DEG terminal that could be used to measure the diagonal
voltage, so the filling factor of the QPC $\nu_{\rm{QPC}}$ is
calculated from the longitudinal four-terminal resistance: In
analogy to magnetotransport through a barrier~\cite{hau88},
$\nu_{\rm{QPC}}$ relates via
$R_{\rm{QPC}}\times{\rm{e^2}}/{\rm{h}}=1/\nu_{\rm{QPC}}+1/\nu_{\rm{Bulk}}$
to the number of occupied Landau levels in the bulk
$\nu_{\rm{Bulk}}=n_{\rm{S}}{\rm{h}}/{\rm{e}}B_{\perp}$. Knowing the
electron density $n_{\rm{S}}$ and Planck's constant $h$, the
measured four-terminal resistance $R_{\rm{QPC}}$ can be directly
converted to $\nu_{\rm{QPC}}$. Figure~\ref{fig:bfield}a) shows data
of QPC3 (split-gate gap $w=250\,\rm{nm}$), measured at a temperature
of $T=0.1\,\rm{K}$ with $B_{\perp}=2\,\rm{T}$.
Plotting $\nu_{\rm{QPC}}$ as a function of $V_{\rm{G1\&G2}}$ reveals
integer filling factors related to the magneto-electric subband
spacing ($\nu_{\rm{QPC}}=2$, $\nu_{\rm{QPC}}=4$) but also smaller
plateaus due to the lifted spin degeneracy ($\nu_{\rm{QPC}}=3$,
$\nu_{\rm{QPC}}=5$). The energy diagram corresponding to
$\nu_{\rm{QPC}}=3$ is shown on the right hand side. The lowest
spin-split plateau is obscured by the 0.7 anomaly~\cite{tho96}. The
same trace repeated at finite source-drain bias
$V_{\rm{SD}}=2\,\rm{mV}$ is shown in Fig.~\ref{fig:bfield}b).
Additional half-plateaus with $\nu_{\rm{QPC}}=2.5$ and
$\nu_{\rm{QPC}}=3.5$ appear which correspond to a situation as
depicted to the right: one spin-split mode is situated in-between
$\mu_{\rm{S}}$ and $\mu_{\rm{D}}$. The 0.7 anomaly has evolved into
a plateau with $\nu_{\rm{QPC}}=1.8$. Figure~\ref{fig:bfield}c) shows
a false color plot of the transconductance
$G_{\rm{TC}}=dG/dV_{\rm{G1\&G2}}$, plotted as a function of
$V_{\rm{SD}}$ and $V_{\rm{G1\&G2}}$. Similar to the transconductance
plots in Figs.~\ref{fig:nonlinear}, regions of integer filling
factor appear as black diamonds. Due to the different sizes of the
subband-split levels (labeled 2, 4 and 6) and the spin-split levels
(3 and 5), the half-plateaus at finite $V_{\rm{SD}}$ appear as black
stripes. A clear deviation from the regular even-odd pattern is
observed at low filling factors, where $\nu_{\rm{QPC}}=0.5,1,1.5$
are replaced by a $\nu_{\rm{QPC}}=1.8$ plateau related to the 0.7
anomaly.

From the extent of the spin-split plateaus the exchange-enhanced
$g$-factor $g^{\ast}$ can be extracted via
$V_{\rm{SD}}=\Delta_{\rm{Spin}}=g^{\ast}\mu_{\rm{B}}B_{\perp}$,
where $g^{\ast}$ is the effective $g$-factor and $\mu_{\rm{B}}$
denotes the Bohr magneton. Compared to the bare $g$-factor of GaAs
($g=-0.44$), we find a strongly enhanced  $g^{\ast}=4.4$ at
$\nu_{\rm{QPC}}=3$ and $g^{\ast}=3.8$ at $\nu_{\rm{QPC}}=5$. Similar
to findings of Thomas at al.~\cite{tho96}, $g^{\ast}$ increases with
lower mode number. However, the magnitude of the exchange
enhancement is different: $0.4<g^{\ast}<1.3$ was reported by Thomas
at al., while our results indicate a much stronger enhancement.
Assuming that disorder reduces the effectiveness of the exchange
enhancement, the observation of strongly enhanced $g$-factors can be
interpreted as another manifestation of the good sample quality.

\section{Conclusion}
In conclusion, we investigated the transport properties of two
differently shaped constrictions that were defined within a
high-mobility 2DEG. Transport spectroscopy in the linear response
regime demonstrates that conductance quantization is observed and
that no scattering centers are found when shifting QPC1 between the
gates. The 0.7-anomaly is investigated by varying the temperature
and by applying a magnetic field perpendicular to the 2DEG.
Depending on the ratio of these two parameters we observe either a
weakening or an enhancement of the 0.7-anomaly which is discussed in
view of spin-polarized edge channels mimicking the 0.7 scenario.
Measurements at finite source-drain bias give access to the subband
spacings of the QPC and reveal that QPC1 is described best by a
short constriction whereas QPC2 shows quantum wire-like
characteristics. In addition to QPCs defined in standard 2DEGs,
resonances at large bias are observed in QPC1 that correspond to
higher order transport conditions. These higher order resonance
conditions give valuable information about the shape of the
confinement potential and enable the reconstruction of the full
saddle point potential in the constriction. Knowing the shape (and
hence the slope) might prove important for the investigation of many
body states like the 0.7 anomaly or the transmission of fractional
quantum Hall states. At finite magnetic field and lower temperature,
a strongly exchange-enhanced $g$-factor is observed which we
interpret as the result of a very smooth confinement potential.
While the measurements on QPCs are possible with high signal to
noise ratio if the gate voltages are always swept in the same regime
and direction, it was not possible to form a stable quantum dot with
the same technique on the same wafer. For future interferometer
experiments on high-mobility samples in the fractional quantum Hall
regime it is desirable to prepare split-gate electrodes on
high-mobility wafers that do not rely on screening electrons at the
X-valley and thereby allow stable gate operation.

\section{Acknowledgments}
We acknowledge the support of the ETH FIRST
laboratory and financial support of the Swiss Science Foundation
(Schweizerischer Nationalfonds, NCCR Nanoscience).

\end{document}